\documentstyle[graphicx]{mn}

\begin{document}
\def\lsun{{\rm L_{\odot}}} 
\def\msun{{\rm M_{\odot}}} 
\def\rsun{{\rm R_{\odot}}} 
\def\RL{R_{\rm L}} 
\title{The Minimum Orbital Period in Thermal--Timescale Mass Transfer} 
\author[A.R.~King et al.]{A.R.~King$^1$, K.~Schenker$^1$, U.~Kolb$^2$,
M.B.~Davies$^1$\\ 
$^1$ Department of Physics \& Astronomy,
University of Leicester, Leicester, LE1 7RH\\
$^2$ Department of Physics \& Astronomy, The Open University, Walton Hall,
Milton Keynes, MK7~6AA}
\maketitle

\begin{abstract}
We show that the usual picture of supersoft X--ray binary evolution as
driven by conservative thermal--timescale mass transfer cannot explain
the short orbital periods of RX~J0537.7--7034 (3.5~hr) and
1E\,0035.4-7230 (4.1~hr). Non--conservative evolution may produce such
periods, but requires very significant mass loss, and is highly
constrained. 
\end{abstract}

\begin{keywords}
stars: evolution --- X--rays: stars --- binaries: close --- 
stars: individual: RX~J0537.7--7034 --- stars: individual: 1E\,0035.4-7230
\end{keywords}

\section{Introduction}
The importance of episodes of mass transfer in a semidetached binary
on a thermal timescale has recently been 
emphasized
in a number of contexts. Such episodes can arise in either of two
ways: 

(i) the donor star fills its Roche lobe while already undergoing
thermal expansion across the Hertzsprung gap, or

(ii) the donor fills its Roche lobe with a mass ratio $q = M_2/M_1$
($M_2$ denotes the donor mass, $M_1$ the accretor mass)
large enough that the Roche lobe radius $\RL$ shrinks on mass transfer
more rapidly than the thermal equilibrium radius $R_{\rm te}(M_2)$
(where this exists)
appropriate to the donor's current mass $M_2$. 

\begin{figure}
 \begin{center}
  \centerline{\includegraphics[clip,width=0.95\linewidth]{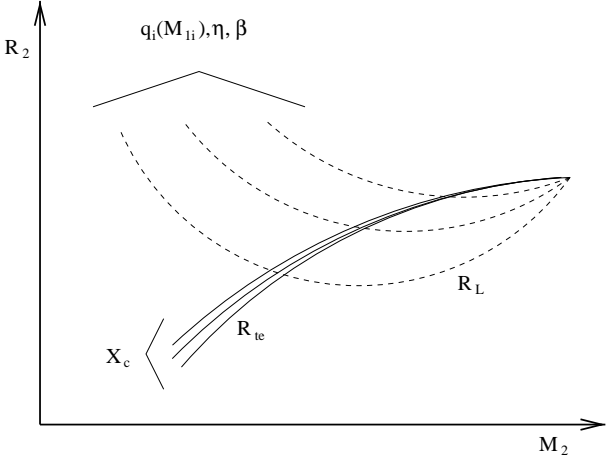}}
  \caption{Schematic diagram with $R_{\rm te}$ (full line) and $\RL$
(broken line) versus $M_2$ 
for Case A mass transfer, after Giannone, Kohl \& Weigert (1968).
While the Roche lobe curve
varies significantly with initial WD mass $M_{1i}$ and
the specification of mass transfer ($\eta$, $\beta$), $R_{\rm te}$
depends only very little
on the evolutionary state along the main sequence (indicated by the
central hydrogen content $X_c$). Thermal--timescale mass transfer may
occur until the two curves intersect 
which happens at different
$M_2$ depending on a particular choice of $\RL$.}
  \label{fig:rrm2}
 \end{center}
\end{figure}

In case (i) the thermal--timescale episode will end once the donor
attains a new thermal equilibrium radius, e.g. at some point after
reaching the Hayashi line. In case (ii) the donor is continually
trying to expand thermally beyond $\RL$ in order to reach $R_{\rm
te}$, so mass is transferred on a thermal timescale while this
condition holds.  Generally this case involves shrinkage of the orbit
to some minimum Roche lobe size, followed by orbital expansion as the
mass ratio $q$ reverses. The thermal--timescale episode ends only when
$R_{\rm te} < \RL$ 
(cf.\ Fig.~\ref{fig:rrm2}; see also van den Heuvel 1992 and references therein).  
Thereafter normal mass transfer continues driven either by systemic
angular momentum losses (descreasing $\RL$) or nuclear evolution of
the donor (increasing $R_{\rm te}$). 

Case (i) arises in the formation of intermediate--mass X--ray binaries
with black--hole accretors, such as GRO~J1655--40 (Kolb et al., 1997;
Kolb, 1998). (In similar systems with neutron--star accretors, such as
Cyg X--2, the condition for case (ii) may hold simultaneously -- King
\& Ritter, 1999; Podsiadlowski \& Rappaport, 2000, Kolb et al.,
2000; 
Tauris et al., 2000). 
In this paper we are mainly concerned with case (ii). This has 
received most attention in connection with the supersoft X--ray
binaries (van den Heuvel et al., 1992). Thermal--timescale mass
transfer from a donor initially on or close to the main sequence, on
to a white dwarf accretor, offers a way of driving accretion rates
$\dot M_1 \sim 10^{-7}\msun\ {\rm yr}^{-1}$ high enough to allow
steady nuclear burning. As well as potentially explaining the observed
supersoft systems, this process allows the white dwarf mass $M_1$ to
grow. If one can arrange that $M_1$ reaches the Chandrasekhar mass $M_C
\simeq 1.44\msun$ this suggests a way of making Type Ia
supernovae. However the difficulty of computing mass transfer on these
timescales meant that early studies of this process simply used the
assumption that mass transfer occurred on a thermal timescale, and
were therefore unable to predict the evolution of the binary
parameters (masses, period, mass transfer rate). Detailed calculations
of these have only recently begun to emerge (Deutschmann, 1998) but
are not yet exhaustive.

Discussion of the evolution of supersoft X--ray binaries would be
greatly eased if observation provided reliable masses. However this is
very difficult for several reasons. For example Greiner et al.\ (2000)
suggest rather low masses $M_1 \simeq 0.6\msun, M_2 \simeq 0.35\msun$
for the short--period system RX~J0537.7--7034 we shall discuss
extensively in this paper. But these and similar estimates use the
assumption that the donor is close to the main sequence. By definition
this cannot be true in thermal--timescale mass transfer, since the
star is not in thermal equilibrium; although it may be quite close to
its main--sequence radius, it can also be considerably smaller than
this (cf.~Deutschmann, 1998). Moreover Greiner et al.'s mass function
uses the HeII emission line, and thus may not reflect the dynamical
motion of the accretor.

Since mass information is so hard to come by, a crucial test for the
thermal--timescale model for the supersoft X--ray binaries is provided
by the discovery of systems with fairly short orbital periods. In
particular RX~J0537.7--7034 (Greiner et al., 2000) has a period of
about 3.5~hr, and 1E\,0035.4-7230 (= SMC\,13) (Schmidtke et al., 1996)
has a period of 4.126~hr. Since the initial mass ratio $q_i$ must be
$\ga 1$ for this type of evolution (see below), one expects initial
donor star masses $M_{2i} \ga 1\msun$ for a typical white dwarf
accretor, and thus an initial orbital period $P_i \ga 10$~hr (see
Section 3 and Fig.~\ref{fig:mpm1}). Evidently considerable orbital shrinkage would
be needed for such systems to reach the periods of RX~J0537.7--7034 and
1E\,0035.4-7230. This is unlikely if the mass transfer is
conservative, as we shall show. However it is probable that much of
the transferred mass is not accreted by the white dwarf, but blown
away from it as a wind (e.g. Li \& van den Heuvel, 1997), allowing
greater orbital shrinkage. Mass loss induced in some way by the mass
transfer process is probably the only way of significantly increasing
the orbital shrinkage, as other angular momentum loss processes such
as magnetic braking or gravitational radiation generally take place on
timescales far longer than the $10^7$~yr characteristic of mass
transfer in the supersoft X--ray binaries.

In this paper we consider thermal--timescale mass transfer and
investigate how much mass loss from the accretor is required if
RX~J0537.7--7034 and 1E\,0035.4-7230 are products of the standard
picture of the supersoft binaries. Our method is to compute the
minimum orbital period analytically for specified rates of mass and
angular momentum loss from the binary. 

\section{Orbital Evolution}
We consider the orbital evolution of a semidetached binary in which a
fixed fraction $1-\eta$ of the mass transferred from the donor (star
2) is lost from the accretor (star 1) with $\beta$ times the specific
angular momentum of the latter. (The quantity $1-\eta$ is called
$\alpha$ by King \& Kolb, 1995 in their general treatment of such
`consequential angular momentum loss' [CAML] mechanisms; the use of
$\eta$ allows more compact formulae in what follows.)  Following the
general method of van Teeseling \& King (1998), setting their
quantities $\dot M_{w2} = \dot J_{\rm sys} = 0, \beta_1 = \beta$, we
find
\begin{equation}
\dot M_{w1} = (1-\eta)\dot M_2,
\label{eta}
\end{equation}
so that 
\begin{equation}
\dot M_1 = -\dot M_2 + \dot M_{w1} = -\eta \dot M_2.
\label{m}
\end{equation}
These definitions give
\begin{equation}
\frac{\dot J}{J} = ( 1 - \eta ) \dot{M}_2 \beta \frac{M_2}{M_1 M},
\end{equation}
with $M = M_1+M_2$. Kepler's 3rd law links angular momentum $J$ and
period $P$ of the orbit as 
\begin{equation}
J^3 = {\rm G}^2 \frac{M_1^3 M_2^3}{M} \frac{P}{2 \pi}.
\end{equation}
The evolution of the orbital period then obeys
\begin{equation}
\frac{\dot P}{P} = -\frac{3\dot M_2}{M_2} +{3\dot M_2\over M_1}
-{2(1-\eta)\dot M_2\over M} +3(1-\eta)(\beta - 1){M_2\dot M_2\over
MM_1}.
\label{rldot}
\end{equation}
From (\ref{m}) we have $M_1 = C-\eta M_2, M
= C + (1-\eta)M_2$, with $C = M_{1i} + \eta M_{2i}$, where $M_{1i},
M_{2i}$ are the initial values of the two masses. Thus (\ref{rldot})
integrates to
\begin{equation}
P \propto M_2^{-3}M_1^{-3(1-\beta)-3\beta/\eta}M^{1-3\beta}
\label{p0}
\end{equation}
for $\eta \neq 0, 1$, the corresponding expression for $\eta =0$
being given by the limit of this expression as $\eta \rightarrow 0$,
while we get the well--known result 
\begin{equation}
P \propto M_2^{-3}M_1^{-3},
\label{p1}
\end{equation}
for conservative mass transfer, i.e. $\eta = 1$.
In terms of the mass ratio $q = M_2/M_1$ we can express the relation
between $M_2$ and $M_1$ as
\begin{equation}
qM_1 = M_{2i} + {1\over \eta}(M_{1i} - M_1),
\label{q}
\end{equation}
leading to
\begin{equation}
{M_{1i}\over M_1} = {q\eta + 1\over q_i\eta + 1}.
\label{m1}
\end{equation}
Thus with $M = (1+q)M_1$, etc, (\ref{p0}) gives finally
\begin{equation}
{P\over P_i} = \biggl({q_i\over q}\biggr)^3\biggl({1+q_i\over
1+q}\biggr)^{(3\beta - 1)} \biggl({1+q\eta\over 1+q_i\eta}\biggr)^{5+
3\beta/\eta},
\label{p}
\end{equation}
where $P_i$ is the initial period.

As a check we note that if all the transferred mass is blown away from
star 1, with the specific angular momentum of that star, we have
$\eta = 0, \beta = 1$; we take the limit of (\ref{p}) as $\eta
\rightarrow 0$ by noting that $(1+q\eta)^{3/\eta} =
\exp[(3/\eta)\ln(1+ q\eta)]$ and using l'H\^opital's rule on the
exponent to give the limit as $\eta \rightarrow 0$ as $e^{3q}$. We
then get 
\begin{equation}
{P\over P_i} = \biggl({q_i\over q}\biggr)^3\biggl({1+q_i\over
1+q}\biggr)^2e^{3(q-q_i)},
\label{w}
\end{equation}
or 
$P\propto M_2^{-3}M^{-2}\exp[3M_2/M_1]$, as found by e.g. King \& Ritter
(1999) for this case.

We are interested in the minimum orbital period attained during the
binary evolution. Regarding $P$ in (\ref{p}) as a function of $q$,
i.e. with $P_i, q_i$ fixed, we find
\begin{equation}
{1\over P}{\partial P\over \partial q} = -{3\over q} - {3\beta - 1\over 1+q} +
{3\beta + 5\eta\over 1+q\eta},
\label{dp}
\end{equation}
so that $P$ is an extremum at
\begin{equation}
q = q_m = {(1-\eta) + \sqrt{(1-\eta)^2 + 9(\eta + \beta -
\eta\beta)}\over 3(\eta + \beta - \eta\beta)} 
\label{qm}
\end{equation}
From (\ref{dp}) we can show that 
\begin{equation}
{1\over P}{\partial^2P\over\partial q^2} = {3\over q^2(1+q\eta)} +
{(3\beta - 1)(1-\eta)\over (1+q)^2(1+q\eta)} 
\label{ddp}
\end{equation}
at $q=q_m$, so $P$ has a minimum there, assuming
$0 \leq \eta \leq 1$.
Hence for given $P_i, q_i$, the smallest orbital period
which can be attained in evolution with mass loss is
given by (\ref{p}) with $q=q_m$. For a given initial period, we still
have the freedom to vary the initial mass ratio $q_i$. 
(This
corresponds to the fact that the period $P_i$ essentially determines
the donor's mean density, almost independently of the primary mass. 
So a donor of given initial mass $M_{2i}$ may initiate mass transfer
at various $P_i$, depending on its evolutionary state, with different
$q_i$, determined by $M_{1i}$.) 
For fixed $P_i$, 
the minimum value of $P$ 
clearly has a {\it maximum}, regarded as a function of
$q_i$, at $q_i = q_m$ 
(see also Fig.~\ref{fig:pmpi_con} which shows the special case $\beta = 1$). 
In other words, ${\rm d} (P(q=q_m))/({\rm d} q_i)_{[P_i {\rm fixed}]} < 0$.
Since $q$ decreases from its initial value $q_i$ through
the evolution this means that given an initial period $P_i$, the
smallest value of the minimum period 
is given by the largest possible value $q_{il}$ of $q_i$. Hence given
an initial period $P_i$, the minimum possible orbital period
$P_m(P_i)$ is given by
\begin{equation}
{P_m\over P_i} = 
\biggl({q_{il}\over q_m}\biggr)^3\biggl({1+q_{il}\over
1+q_m}\biggr)^{3\beta -1}
\biggl({1+q_m\eta\over 1+q_{il}\eta}\biggr)^{5+ 3\beta/\eta}, 
\label{pm}
\end{equation}
with $q_m$ given by (\ref{qm}), and $q_{il}$ the largest possible
value of the initial mass ratio $q_i$.
Although  $\beta$ is of course irrelevant in the conservative case
($\eta=1$), larger angular momentum losses in the wind allow even smaller
ratios $P_m/P_i$ for the same $q_i$.

It is likely that thermal--timescale mass transfer is still
going on at the minimum orbital period, since by Roche geometry we
have
\begin{equation}
\RL \propto f(q)M^{1/3}P^{2/3}.
\label{prl}
\end{equation}
where $f(q)$ is a slowly increasing function of $q$ 
given e.g.\ by the approximation from Eggleton (1983).
Thus
\begin{equation}
{\dot\RL\over\RL} = {\dot q \over f}{{\rm d}f\over {\rm d}q} +
{\dot M\over3M} + {2\dot P\over 3P},
\label{dot}
\end{equation}
which tells us immediately that the period reaches its minimum value
at larger mass ratios than $\RL$ as $\dot q < 0, \dot M \leq 0$.
In general $\RL$ initially shrinks much more rapidly than $R_{\rm te}$.
Thus the condition $R = R_{\rm te}$ signalling the end of
thermal--timescale mass transfer is reached only once $\RL$ is
increasing after passing through its minimum value,
i.e.\ thermal timescale mass transfer 
usually 
stops only {\it after} the minimum orbital period is reached. 
However it is conceivable that in some cases $R_{\rm te}$ shrinks very
rapidly on mass loss, and individual cases must be checked. For the
purposes of this paper we note than the true minimum period for
thermal--timescale could if anything be longer than the values of
$P_m$ we find here.

In what follows we will assume that $\beta = 1$, i.e. that the mass
lost from the accretor has the same specific angular momentum as this
star. This for example includes any form of mass loss from orbits with
circular symmetry about the accretor, as might occur from an accretion
disc. In this case (\ref{pm}) simplifies to  
\begin{equation}
{P_m\over P_i} = 
\biggl({q_{il}\over q_m}\biggr)^3\biggl({1+q_{il}\over 1+q_m}\biggr)^2
\biggl({1+q_m\eta\over 1+q_{il}\eta}\biggr)^{5+ 3/\eta}, 
\label{pm0}
\end{equation}
with 
$1 \leq q_m \leq 1.387$ now given by
\begin{equation}
q_m = {(1-\eta) + \sqrt{(1-\eta)^2 + 9}\over 3}.
\label{qm0}
\end{equation}
and $q_{il}$ the largest possible value of the initial mass ratio
$q_i$. Figure \ref{fig:pmpi_con} shows the ratio $P_m/P_i$ as a function of 
$q_i$ 
for the two extreme cases $\eta = 1, 0$.

\section{Minimum Periods}

\begin{figure}
 \begin{center}
  \centerline{\includegraphics[clip,width=0.95\linewidth]{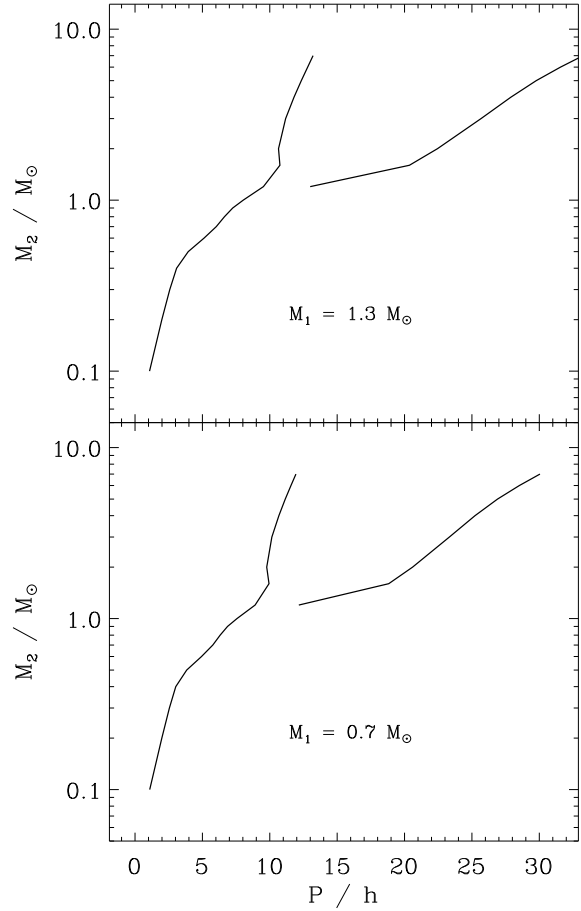}}
  \caption{Typical range of $P_i$ versus $M_{2i}$ for different WD
masses, where the lines indicate the main sequence boundaries. For
initial donor star masses in the expected range between $1 \dots 5
\msun$ the short period edge (= ZAMS) lies at about 10~hr, and is fairly
constant for $M_{2i} \protect\ga 1.3 \msun$. The additional line marks the
terminal main sequence (TMS), more precisely the point of minimum
$T_{\rm eff}$ 
(where applicable).}
  \label{fig:mpm1}
 \end{center}
\end{figure}

The work of the last Section, especially eqn (\ref{pm}), shows that to
compute the ratio $P_m/ P_i$ of minimum to initial period we need to
specify the largest possible value $q_{il}$ of the initial mass ratio.
For a white--dwarf accretor
the short observed periods of RX~J0537.7--7034 and 1E\,0035.4-7230 then
constrain the mass--loss parameter $\eta$. Clearly to reach such short
periods it is preferable to start from the shortest initial periods
$P_i$, which in turn are given by assuming that the donor is still
very close to the ZAMS when it initiates mass transfer. We can easily
show that the initial donor mass $M_{2i}$ must be $\ga 1.9\msun$ in
order to explain the 3.5~hr period of RX~J0537.7--7034, by iterating
the formula (\ref{pm0}), using the fact that we must clearly have 
$q_{il} > q_m$. 
For the lowest--mass white dwarf we consider ($M_{1i} = 0.7\msun$)
this implies $M_{2i} > 0.7q_m\msun$. Thus in the conservative case
$\eta=1$ we have $q_m=1, M_{2i} > 0.7\msun$, and Fig.~\ref{fig:mpm1}
shows that $P_i \ga 6$~hr for a ZAMS star of this mass. We can now use
Fig.~\ref{fig:pmpi_con} with the restriction $P_m/P_i < 3.6/6 = 0.58$
to find $q_{il} > 2.36$ and thus $M_{2i} > 1.66\msun$. Figure
\ref{fig:mpm1} now shows that $P_i \simeq 10$~hr, and we may iterate
using Fig.~\ref{fig:pmpi_con} to get $q_{il} > 3.38, M_{2i} >
2.37\msun$. Further iteration fails, as Figure \ref{fig:mpm1} shows
that the new higher estimate for $M_{2i}$ does not increase the
estimate for $P_i$.  In a similar way we find $q_{il} > 2.7, M_{2i} >
1.89\msun$ for the other extreme case $\eta = 0$.

This argument essentially fixes the minimum value of $P_i$ at about
10~hr (see Figure \ref{fig:mpm1}), so the orbital shrinkage
required to explain the 3.5~hr period of RX~J0537.7--7034 has 
$P_m/P_i < 0.35$. 

Given an initial white dwarf mass $M_{1i}$ and mass
ratio $q_{il}$ we can predict the period ratio $P_m/P_i$ for any given
$\eta$. The initial mass ratio $q_{il}$ must obey two constraints
(added as vertical lines in Fig.~\ref{fig:pmpi_con}):

\begin{figure}
 \begin{center}
  \centerline{\includegraphics[clip,width=0.95\linewidth]{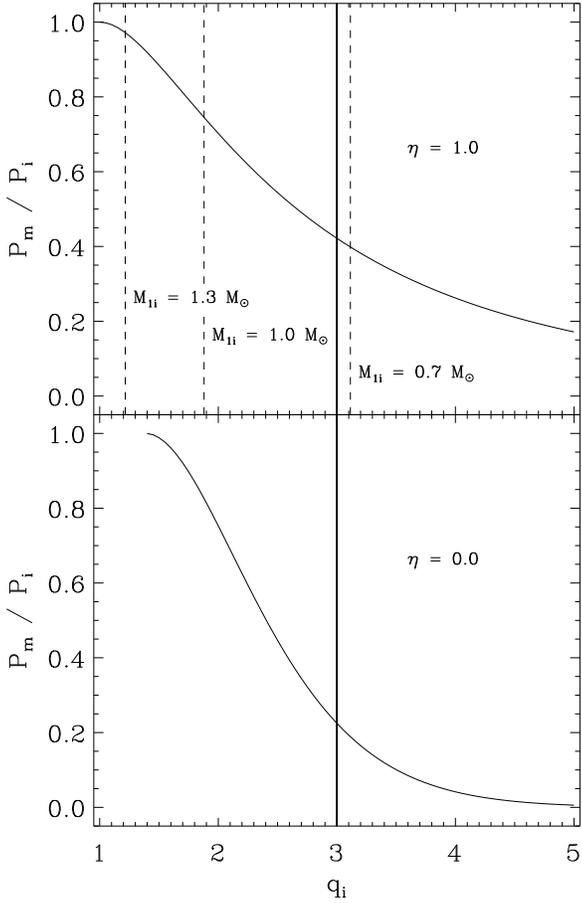}}
  \caption{$P_m/P_i$ as a function of $q_i$ for $\eta = 1, 0$
according to (\ref{pm0}).
Additional constraints for various initial WD masses
resulting from the Chandrasekhar limit are indicated by dashed lines in the
conservative case. 
They mark the critical values of $q_i$ where the WD
mass has grown to $M_C$ precisely upon reaching $P_m$.
An upper limit at $q_i = 3.0$ due to delayed dynamical instability
(DDI) is also indicated.} 
  \label{fig:pmpi_con}
 \end{center}
\end{figure}       

(i) the system must avoid the `delayed dynamical instability' 
(DDI; see Webbink, 1977; Hjellming, 1989), which occurs when 
sustained thermal--timescale mass transfer 
exposes inner layers with a flat entropy gradient, 
and

(ii) the white dwarf mass cannot exceed $M_C$ before the minimum
period $P_m$ is reached.

In practice the first of these constraints requires 
$q_{il} < q_{\rm DDI} \simeq 3$. 
Hjellming (1989) found this value for a donor with $M_{2i} = 3 \msun$
near the terminal main sequence (TMS), whereas Kalogera \& Webbink (1996)
seem to prefer even smaller limiting ratios around $2.5$.
Kolb et al.\ (2000) used Mazzitelli's stellar code to calculate a test
sequence with constant primary mass $0.75 \msun$ where mass transfer
starts from a $3 \msun$ near--TMS star. A DDI occured at donor mass
$2.6 \msun$, in perfect agreement with Hjellming's prediction.  
The critical maximum mass ratio $q_{\rm DDI}$ depends on the stellar
structure, and therefore on the stellar input physics. In particular,
$q_{\rm DDI}$ is probably sensitive to the degree of convective 
core overshooting, as this determines the size of the convective
core. This is highlighted by the fact that Kolb et al.\ (2000) find
$q_{\rm DDI} = 2.9$ for early massive case B mass transfer, while
Tauris et al.\ (2000) find $q_{\rm DDI} = 3.6$, using an  
updated Eggleton code. 

The second constraint 
is quite severe for large initial white dwarf masses and for 
$\eta \simeq 1$. In particular, for $M_{1i} = 0.7\msun$ it requires 
$q_{il} < 3.11$, in mild contradiction with the requirement 
$q_{il} > 3.38$ we found above.

\begin{figure}
 \begin{center}
  \centerline{\includegraphics[clip,width=0.95\linewidth]{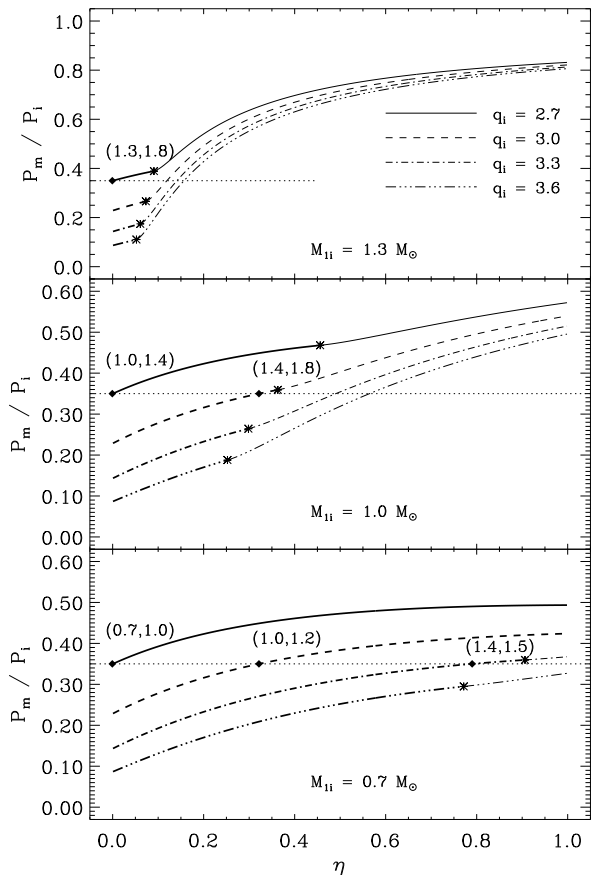}}
  \caption{$P_m/P_i$ versus $\eta$ for various $q_{i}$ as labelled.
Only the part of each curve in thick linestyle reaches $P_m$ before
the WD grows to the Chandrasekhar mass and a SN Ia occurs, with an
  asterisk marking the critical $\eta$. Beyond that the current period
  at the SN event has been given as the smallest possible fraction
  $P_m/P_i$ during the thermal--timescale mass transfer phase. The
  horizontal dotted line at $P_m/P_i = 
0.35$ marks the fraction {\em at least} required to create a system
like RX~J0537.7--7034 starting from the ZAMS with $P_i \approx 10$~hr.
The numbers in brackets show pairs of masses $(M_1(q_m),M_2(q_m))$
each belonging to a set of parameters ($\eta,q_i$) just fulfilling
this requirement and marked by a diamond. DDI constraints (that would
exclude the two larger initial mass ratios) have not been considered
in this graph.} 
  \label{fig:pmpi}
 \end{center}
\end{figure}       

Figure \ref{fig:pmpi} shows $P_m/P_i$ versus $\eta$ for various values
of $q_{il}$ The stars denote combinations $M_{1i}, q_{il}, \eta$ where
$M_1$ reaches $M_C$ precisely at $q=q_m$. To the right of these
positions we take $P_m$ as the minimum period actually achieved,
i.e.\ the period where the white dwarf reaches $M_C$.

Each panel of this Figure shows:

1. The horizontal line $P_m/P_i = 0.35$, 
   i.e.\ the upper limit required by RX~J0537.7--7034.

2. Only those curves which actually manage to cross this line for
   $\eta \geq 0$. 
   In particular we plot the curve for the limiting
   value of $q_i$ such that the curve just crosses $P_m/P_i = 0.35$
   at $\eta = 0$. This gives a lower limit on $q_i$ which is the same
   for each mass of the WD.

3. At some of the extreme solutions, i.e. wherever the line
   $P_m/P_i = 0.35$ is crossed before a SN occurs,
   the current values of $M_1, M_2$ are given. Note that with growing
   $M_{1i}$ the donor mass has to go up as well for the same $q_i$.
   Furthermore $M_1$ at the minimum period comes closer and closer to 
   the Chandrasekhar mass as we increase $\eta$.

\section{Discussion}
Figure \ref{fig:pmpi} shows already that the short orbital period of
RX~J0537.7--7034 poses very severe constraints if this system results
from thermal--timescale mass transfer. In particular

(a) Conservative evolution ($\eta = 1$) is possible only for initial
mass ratios $q_i > 3$ which probably make the system vulnerable to the
delayed dynamical instability. Full evolutionary calculations are
required to check if there are any evolutionary tracks which avoid
it. If such tracks exist, the initial white dwarf mass in
RX~J0537.7--7034 must have been low ($\la 0.7\msun$), but the current
system masses must be fairly high (e.g. $q_i = 4$ requires $M_1 =
1.17\msun, M_2 = 2.33\msun$).  These are of course in conflict with
the mass estimates of Greiner et al. (2000), but this may not by
itself be fatal (see the remarks in the Introduction).

\begin{figure}
 \begin{center}
  \centerline{\includegraphics[clip,width=0.95\linewidth]{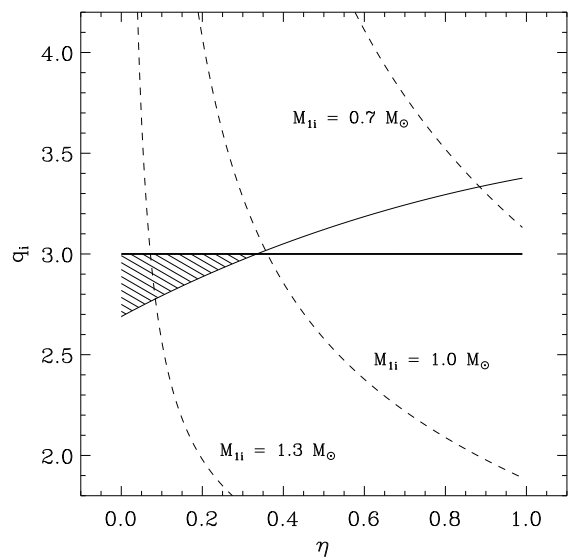}}
  \caption{Borderline in the plane of $\eta$ and $q_i$ needed to obtain
    $P_m/P_i = 0.35$. The allowed range is limited to the small area
above the curved line and below the horizontal DDI limit at $q_i = 3.0$.
Additional constraints are shown for various $M_{1i}$ (dashed lines).
The upper limits to $q_i(\eta)$ (or $\eta(q_i)$) result from
the WD growing beyond its Chandrasekhar mass before reaching $P_m$,
see Fig.~\ref{fig:pmpi} and text for details. Initially massive WDs
with $\protect\ga 1.0\msun$ restrict the allowed range to 
$\eta \protect\la 0.3$ independently of DDI.}
  \label{fig:eta_q_rel}
 \end{center}
\end{figure}       

(b) Non--conservative evolution ($\eta < 1$) does allow tracks
with $q_i \leq 3$ which probably avoid the delayed dynamical
instability. However the ranges of $\eta$ and $q_i$ are still very
tightly constrained. 
Figure \ref{fig:pmpi} shows that $2.7 \la q_i \la 3$, 
$0 < \eta \la 0.3$, with $\eta$
and $q_i$ correlated as in Figure \ref{fig:eta_q_rel}.
For $M_{1i} \ga 1.0\msun$ $\eta$ has to be even lower, and generally
upper limits on $q_i$ are given for each $M_{1i}$ in addition to DDI.
Note that for the largest values of $\eta$ the white dwarf in
RX~J0537.7--7034 is predicted to be close to the Chandrasekhar limit
unless the DDI limit imposes a more severe restriction than the lowest
possible WD mass (as is the case in Fig.~\ref{fig:eta_q_rel} for $q_i
\la 3, M_{1i} \ga 0.7\msun$).

Even given these tight constraints, the evolutions discussed above
require the system to have come into contact with the donor still very
close to the ZAMS. As orbital angular momentum losses (e.g. via
magnetic braking) are probably negligible for the likely initial
masses $M_{2i} \ga 2\msun$, this also requires the initial orbital
separation to lie in an extremely narrow range. Evidently to make a
system like RX~J0537.7--7034 or 1E\,0035.4-7230 by the
thermal--timescale route is a very rare event. In line with this,
Deutschmann (1998) found no orbital periods shorter than about 6~hr in
his detailed calculations
with solar metallicity.
Furthermore these requirements have been derived assuming that
RX~J0537.7--7034 and 1E\,0035.4-7230 are just at the minimum period
$P_m$, so the conditions might be even harder to meet.
The probability of observing such a system is larger near the end of the
thermally unstable phase because the mass transfer rate 
decreases (in both Deutschmann's and our own full computations).

We have so far neglected the effect of tidal interactions on the
orbital evolution of the binary. Tauris \& Savonije (2000) show that
these can be important in low--mass X--ray binaries by translating
spin angular momentum losses (via e.g. magnetic stellar wind braking)
into orbital losses when tidal synchronization occurs. However for the
binaries we consider here, any effect before the beginning of mass
transfer simply allows a shorter $P_i$ for a wider range of
systems. Once synchronism is achieved, the angular momentum of a
lobe--filling donor is less than about 10 per cent of the orbital
angular momentum (because the gyration radii of typical donor stars is
small, $r_{\rm g}^2 \leq 0.10 {\ldots} 0.20$). Even transferring all
of this to the orbit in the course of the evolution would lead to only
a marginal shift towards longer minimum periods, leaving our
conclusions unchanged.

The work of this paper suggests that while the thermal--timescale mass
transfer model for the supersoft X--ray binaries has many desirable
features, it may not be possible to use it to describe all of the
supersoft binaries, as well as SNe Ia progenitors. For example, if
highly non--conservative mass transfer is as common as seems to be
required to explain RX~J0537.7--7034 or 1E\,0035.4-7230, this would
make building up the white dwarf mass to produce a Type Ia supernova
highly problematical.  It may therefore be necessary to consider other
possibilities (e.g. van Teeseling \& King, 1998; King \& van
Teeseling, 1998).

\section{Acknowledgment}
ARK gratefully acknowledges the support of a PPARC Senior Fellowship.
This work was partially supported by a PPARC Rolling Grant for
Theoretical Astrophysics.

\end{document}